\documentstyle[aps]{revtex}

\begin{document}
\title{ Restoration and Dynamical Breakdown of the $\phi\to -\phi$ Symmetry
        in the (1+1)-dimensional Massive sine-Gordon Field Theory}
\author{Wen-Fa Lu }
\address{CCAST(World Laboratory) P.O. Box 8730, Bejing, 100080 \\
         and \\
         Department of Applied Physics, Shanghai Jiao Tong University,
         Shanghai 200030, China
         \thanks{mailing address,E-mail: wenfalu@online.sh.cn}}
\maketitle
\begin{abstract}
Within the framework of the Gaussian wave-functional approach, we investigate
the influences of quantum and finite-temperature effects on the $Z_2$-symmetry$
(\phi \to -\phi)$ of the (1+1)-dimensional massive sine-Gordon field theory. It
is explicitly demonstrated that by quantum effects the $Z_2$-symmetry can be
restored in one region of the parameter space and dynamically spontaneously
broken in another region. Moreover, a finite-temperature effect can further
restore the $Z_2$-symmetry only.
\end{abstract}
\vspace{24pt}

\section{Introduction}
\label{1}

The massive sine-Gordon field theory (MSGFT) \cite{1} is an interesting and
important one in quantum field theory, particle physics and condensed matter
physics. It is a simple generalization of the massless sine-Gordon field theory
(SGFT) \cite{2} which received an extensive investigations \cite{3,4,5},
with a vacuum angle $\theta$ added in the argument of the cosine and a mass
term $m_0^2\phi^2$ added in the Lagrangian. Nevertheless, different from the
SGFT, the Lagrangian of MSGFT is no longer invariant under the transformation
of the field $\phi\to (\phi+{\frac {2n\pi}{\beta}})$, and at the classical
level, the symmetry $\phi \to - \phi$ ($Z_2$-symmetry) can be spontaneously
broken, which is an important phenomenon for quantum field theory and particle
physics \cite{6}. Furthermore, the $1+1$-dimensional (2D) MSGFT is equivalent
to many other important models, such as the massive Schwinger model at a
special coupling strength and on the zero-charge sector \cite{7}, the massive
Schwinger-Thirring model on the zero-charge sector \cite{8,9}, the
two-dimensional (2D) lattice Abelian Higgs model \cite{10},  the 2D neutral
Yukawa gas \cite{11}, and so on. Also, the 2D MSGFT is useful in the
investigations of fluid membranes \cite{12} and vortices in 2D superconductors
\cite{13}. Therefore, it is of general importance to explore various properties
of the MSGFT.

In fact, many investigations of the 2D MSGFT have existed. Early in 1970s, the
2D MSGFT was analyzed within the framework of constructive quantum field theory.
The existence of the 2D MSGFT was proved for a certain region of the parameter
space \cite{1,8,14}, and Fr\"ohlich $et \ al$ also showed the nontriviality of
the scattering, the existence of the one boson states (Fermion-anti-Fermion
bound state), the zero-charge of the physical states in the 2D MSGFT, and so on
\cite{8,9}. Recently, in order to reveal the phase structure of the 2D Abelian
Higgs model \cite{10} and discuss the intrinsic finite size effect in 2D
superconductors \cite{13}, the 2D MSGFT with a finite momentum cutoff was
treated by renormalization-group technique. Additionally, as an equivalent
system of the massive Schwinger model, the MSGFT was investigated for large
$m_0^2$ by Fermi-mass perturbation \cite{7}(1975) \cite{15,16,17}, on the light
front \cite{18} or at finite densities and/or temperatures \cite{19} to discuss
the chiral condensate, quark confinement, the $\theta$-vacuum, particle spectra,
$etc.$. Using the Gaussian wave-functional approach (GWFA) \cite{20}, we also
had a quantitative investigation to the ground, one- and two-particle states of
the 2D MSGFT with zero vacuum angle \cite{21}.

Nevertheless, we feel that for the 2D MSGFT there are still many problems worth
investigating. For instance, the spontaneous breakdown of $Z_2$-symmetry
($Z_2$SSB) is such an interesting phenomenon. As was stated in the first
paragraph, $Z_2$-symmetry of the classical vacuum of the 2D MSGFT is
spontaneously broken \cite{15,16,21}. The $Z_2$SSB of the classical vacuum
means that the classical potential has a minimum, for example, at two points in
the field space $\phi=\pm\hat{\phi}_1$ (here, $\hat{\phi_1}$ is positive), or
say, is two-fold degenerate (infinitely degenerate for the SGFT), and
accordingly the classical vacuum is {\bf occasionally} located at one of the
two points $\phi=\pm\hat{\phi}_1$ and consequently $Z_2$-symmetry is broken.
Similarly, at a quantum level, when effective potential for a quantum field
theory is, for example, two-fold degenerate, the quantum vacuum is spontaneously
located at a non-zero field point, and accordingly $Z_2$-symmetry is also
spontaneously broken. In the 1970's, Ref.~\cite{16} infered that for the 2D
MSGFT the $Z_2$SSB at a classical level would be usually maintained at a
quantum level, and Ref.~\cite{15} pointed out qualitatively, based on a
semiclassical calculation, that for the bosonic equivalent to the massive
Schwinger model $Z_2$-symmetry suffers spontaneous breakdown in some case.
Besides, Ref.~\cite{1} pointed out that the $Z_2$-symmetry in the 2D MSGFT is
presumably dynamically broken in some case. Recently, we also demonstrated
explicitly the existence of the $Z_2$-SSB phenomenon within the framework of
the GWFA \cite{21}. However, as compared to what occurs at the classical level,
the influence of quantum effects on the $Z_2$-SSB phenomenon is not clear.
Usually, quantum effects can alter the $Z_2$-symmetry of the vacuum, and
accordingly turn a symmetric vacuum at the classical level into an asymmetric
one at the quantum level, or $vice \ verse$, an asymmetric vacuum at the
classical level into a symmetric one at the quantum level. The latter is called
the restoration of $Z_2$-symmetry \cite{22,3,6}(1996), and the former is called
the dynamical breakdown of $Z_2$-symmetry ($Z_2$DSB) in this paper. By
dynamical breakdown \footnote{Originally, ``dynamical symmetry breakdown'' was
first used to mean that the chiral symmetry enjoyed by the classical Lagrangian
of the NJL model is broken by quantum radiation corrections, to see
Ref.~\cite{23}.}, we mean that it is just by quantum effects that the
$Z_2$-symmetry enjoyed by the classical vacuum is spontaneously broken at a
quantum level. When $Z_2$DSB occurs, we also say the $Z_2$-symmetry is
dynamically spontaneously broken, which is really involved in the type of the
phenomenon occured in the pure $\lambda\phi^4$ model \cite{24}. For a field
theory, the occurences or disappearances of the $Z_2$-symmetry restoration and
$Z_2$DSB can be seen through a comparison between the two regions of the
parameter space where the $Z_2$SSB occurs respectively at the classical and
quantum levels. If there is a parameter region where the classical vacuum is
asymmetric and the quantum vacuum symmetric, then in that region, the symmetry
breakdown phenomenon is compressed and $Z_2$-symmetry is restored by quantum
effects. If there is a parameter region where the classical vacuum is
symmetrical and the quantum vacuum asymmetrical, then in that region, the
symmetry breakdown phenomenon is enhanced and $Z_2$-symmetry is dynamically
spontaneously broken by quantum effects. Obviously, Fig.1 in Ref.~\cite{21}
suggested the occurence of the $Z_2$-symmetry restoration, but failed to give
such a parameter region where the $Z_2$-symmetry restoration phenomenon occurs.
Furthermore, we donot know if the $Z_2$DSB phenomenon really occurs in the 2D
MSGFT. This paper will address the above problems. Within the framework of the
GWFA, we shall continue to investigate the 2D MSGFT with zero vacuum angle,
explicitly give the region of the parameter space where the $Z_2$-symmetry is
restored, discuss $Z_2$DSB, and analyze the influence of a finite temperature
effect on them.

Both the $Z_2$-symmetry restoration and $Z_2$DSB are interesting and important.
They are the inverse of each other, and can correspond to some phase transitions
which occur at inverse directions. Also they mirror some aspects of quantum
effects. Besides, the $Z_2$-symmetry restoration is the reverse of the fate of
the fase vacuum \cite{25}. Early in 1981, Rajaraman and Lakshmi proposed at the
first time the concept of symmetry restoration and demonstrated explicitly the
restoration of the $Z_2$-symmetry in a special 2D $\phi^6$ model \cite{22},
with the loop-expansion effective-potential method \cite{24}. Later, a
re-investigation of the same model appealing to the same method revealed the
dependence of the symmetry-restoring result upon the renormalization condition
\cite{26}. Later again, a Monte Carlo numerical study about another slightly
different 2D $\phi^6$ theory with three degenerate vacua showed that the
quantum corrections make the vacuum $Z_2$-symmetrical \cite{27}. Recently, we
re-investigated the above two $\phi^6$ models with the GWFA and demonstrated
well the restoration of the $Z_2$symmetry \cite{28}. Although the above two
models can display the restoration of the $Z_2$-symmetry, they (in the
continuum, not the lattice case) are intrinsically dependent upon the
renormalized condition \cite{26,28,29} \footnote{Here, the word
``intrinsically'' means that the dependence upon the renormalized condition is
closely related to the intrinsic nature of those two models themselves that
there are three terms $\phi^2,\phi^4$ and $\phi^6$ in the potential of each of
the above two $\phi^6$ models but there are only two parameters \cite{28,29}.},
and the model in Ref.~\cite{27} is little typical because its classical vacuum
is three-fold degenerate. The $\lambda\phi^4$ model is a good one for displaying
the $Z_2$-symmetry restoration, and was investigated in various dimensions
through the GWFA \cite{30}. In this paper, one will see that the MSGFT is also
a good field-theoretic example for displaying the $Z_2$-symmetry restoration.
As for the dynamical breakdown of the $Z_2$-symmetry, its existence was shown
only in the pure $\lambda\phi^4$ model \cite{24}, and, to our knowledge, no
other investigations demonstrated the occurrence of the $Z_2$DSB in any other
scalar field theories. In this paper, we shall show that the 2D MSGFT also
suffers $Z_2$DSB.

As was mentioned above, for the 2D MSGFT, quantum effects can extend and also
shrink the parameter region where the classical vacuum is $Z_2$-symmetrical.
Then, when a finite temperature is introduced into the 2D MSGFT, will the
parameter regions of the $Z_2$-symmetry restoration and $Z_2$DSB by quantum
effects be extended or shrinked by the finite temperature effect? This is an
interesting problem. When the $Z_2$-symmetry restoration region is extended, or
the $Z_2$DSB region is shrinked, we say the $Z_2$-symmetry is restored by a
finite temperature effect \cite{31}. This phenomenon is important in particle
physics, cosmology and condensed matter physics \cite{31,32,33} \cite{6}(1996).
In the present paper, using the GWFA in thermofield dynamics \cite{34,33,35,36},
we shall consider the influence of a finite temperature on the $Z_2$-symmetry
for the 2D MSGFT with zero vacuum angle. One will see that for the 2D MSGFT, a
finite temperature effect can further restore the $Z_2$-symmetry and would
prevent quantum effects from dynamically breaking the $Z_2$-symmetry.

  For both the zero- and finite-temperature cases, this paper will make use
of the GWFA for investigations. Although it is difficult to control the
accuracy of the GWFA, the GWFA has succeeded in extracting the non-perturbative
information of many field theoretical models \cite{37,21}. So, we feel that for
the $Z_2$-symmetry restoration and $Z_2$DSB phenomena in the 2d MSGFT, the
quantitative result of the GWFA is necessary and useful, at least, can provide
a basis and reference for further investigations.

This paper is organized as follows. Sect.II will discuss the classical vacuum
of the 2D MSGFT so as to lay the foundation for studying the $Z_2$-symmetry
restoraton and $Z_2$DSB phenomena. In Sect.III, we shall give the finite
temperature Gaussian effective potential (FTGEP), the zero-temperature limit of
which is just the Gaussian effective potential (GEP) of quantum field theory.
The restoration and dynamical breakdown of the $Z_2$-symmetry by quantum
effects will be investigated in Sect.IV. In Sect.V we shall consider the
influence of a finite temperature effect on the zero-temperature results, and
Sect. VI will conclude this paper with a brief discussion.

\section{Classical Vacuum}
\label{2}

In this section, we give a discussion about the classical aspects of the 2D
MSGFT for the convenience of latter analyses. Although the analyses in the
present section is simple, and some results are straightforward or were
mentioned in Ref.~\cite{21}, they are the basis of the investigation in this
paper.

For the (1+1)-D MSGFT with zero vacuum angle $\theta=0$, the Lagrangian is
\footnote{Notice that the sign of the term ${\frac {m^2}{\beta^2}}
cos(\beta\phi_x)$ is the plus (+), which is identical with that of Eqs.(I.1)
and (III.1) in Ref.~\cite{8}, but contrary to that in Ref.~\cite{1}.}
\begin{equation}
{\cal L}={\frac {1}{2}}\partial_\mu \phi_x \partial^\mu \phi_x-
                      {\frac {1}{2}}m_0^2\phi_x^2-
          {\frac {m^2}{\beta^2}}(1-\cos(\beta\phi_x)) \equiv
          {\frac {1}{2}}\partial_\mu \phi_x \partial^\mu \phi_x-U(\phi_x),
\end{equation}
with $\phi_x\equiv\phi(x)$, where $m_0$ and $m$ are in mass dimension and the
dimensionless $\beta$ is the coupling parameter. In the case of $m_0=0$, Eq.(1)
describes the SGFT, and when $\beta^2\to 0$, Eq.(1) describes a free theory
with the squared mass ($m_0^2+m^2$)(if $m_0^2+m^2>0$). Because
$|\cos(\beta\phi_x)|\le 1$ (hereafter $| \cdots |$ represents absolute value),
we have to maintain a positive $m_0^2$ in Eq.(1) for avoiding an
unbounded-below vacuum \footnote{We do not know if quantum effects make the 2D
MSGF meaningful for a negative $m_0^2$.}. Similar to SGFT \cite{2}, it is always
viable to have $\beta\ge 0$ because the sign of $\phi$ is free to be redefined.
Nevertheless, different from the SGFT, both the positive and the negative $m^2$
should be considered in Eq.(1) because the physics of the negative is not
equivalent to the one of the positive.

Evidently, the Lagrangian Eq.(1) is invariant under the transformation of
$\phi\to -\phi$. Let us discuss if the $Z_2$-symmetry of the Lagrangian is
enjoyed by the classical vacuum. Consider the extremum condition of the
classical potential $U(\phi)$ (${\frac {d U(\phi)}{d \phi}}=0$)
\begin{equation}
\beta\phi+{\frac {m^2}{m_0^2}} \sin(\beta\phi)=0 \; ,
\end{equation}
and the second derivative
\begin{equation}
{\frac {d^2 U(\phi)}{(d \phi)^2}}=
m_0^2+m^2\cos(\beta\phi) \; .
\end{equation}
For any solution of Eq.(2) $\phi=\hat{\phi}$, the potential $U(\hat{\phi})$ can
be expressed as
\begin{equation}
U(\hat{\phi})={\frac {2 m^2}{\beta^2}}\sin^2({\frac {\beta\hat{\phi}}{2}})
             (1+{\frac {m^2}{m_0^2}} \cos^2({\frac {\beta\hat{\phi}}{2}})) \; .
\end{equation}
Obviously, when $m^2$ is positive, $U(\hat{\phi}=\hat{\phi_1}\not= 0)$ is
greater than the value $U(\hat{\phi}= 0)=0$. When $m^2<0$, Eq.(2) can be solved
with graphics, and Fig.1 solves graphically Eq.(2) with $\beta=3.0$. In Fig.1,
the one,three and seven intersection points of the three lines I,II, and III
with the sine curve are the solutions of Eq.(2) in the cases of
${\frac {m_0^2}{m^2}}=-1.2, -0.5, -0.1$, respectively. From Fig.1, it can be
seen that the more near the negative ratio ${\frac {m_0^2}{m^2}}$ is to zero,
the greater the number of the solutions of Eq.(2) is. In general, for any given
$m_0^2, m^2$ and $\beta$, Eq.(2) can have $(2N+1)$ roots (here, $N$ is a
non-negative integer). When $m^2<0$ and ${\frac {|m^2|}{m_0^2}}<1$, Eq.(2) has
a unique zero root $\hat{\phi}= 0$ because $\beta\phi>\sin(\beta\phi)$ for the
case of $\phi>0$. Thus, both for the case of $m^2>0$ and for the case of $m^2<0$
with $|m^2|<m_0^2$, the classical potential $U(\phi)$ is absolutely minimized
at $\phi=0$, and so the classical vacuum is symmetrical and possesses the same
$Z_2$-symmetry with the Lagrangian. In the other case, $i.e.$, for any negative
$m^2$ with $|m^2|>m_0^2$, the situation is completely different, and there exist
$2N$ non-zero roots of Eq.(2) $\hat{\phi}= \pm\hat{\phi}_i$ ($i=1,2,\cdots,N$)
besides the zero root $\hat{\phi}=0$, and $U(\phi)$ gets to a maximum at $\phi=0$
because ${\frac {d^2 U(\phi)}{(d \phi)^2}}$ of Eq.(3) is negative at $\phi=0$.
Then for the case of $m^2<0$ with $|m^2|>m_0^2$, the classical potential
$U(\phi)$ must be minimized at $\phi=\pm\hat{\phi}_1, \hat{\phi}_3,
\hat{\phi}_5, \cdots$, and $Z_2$-symmetry of the classical vacuum is
spontaneously broken. A further analysis indicates that the classical vacuum is
located at $+\hat{\phi}_1$ or $-\hat{\phi}_1$ which are nearest to the zero
root. The parameter space corresponding to the symmetric and the asymmetric
phases of the classical vacuum is plotted in Fig.2. For any $\beta\not=0$, on
the parameter plane $m_0^2$-$m^2$, the boundary where the symmetric and the
asymmetric vacua coexist corresponds to the ray, from the origin, with the
slope $-1$, in the fourth quadrant of the parameter plane (Fig.2(a)), and the
upper and lower regions of the boundary correspond to the $Z_2$-symmetric and
asymmetric vacua, respectively. For any given $m_0^2\not=0$, on the parameter
plane $m_2$-$\beta^2$, the boundary where the symmetric and the asymmetric
vacua coexist is the line $m^2=-m_0^2$ in the second quadrant of the parameter
plane (Fig.2(b)), and the right region of the boundary corresponds to symmetric
vacua, whereas the left region to asymmetric vacua. As an explicit illustration,
the classical potentials at $\beta=3.0$ and $m_0^2=0.6$ are depicted in Fig.3
for the cases of $m^2=6.0, -0.5, -3.0$, and $-6.0$, respectively. (In the
numerical computation, unit mass is regarded as 1 and hence quantities concerned
can be taken as dimensionless.)

When the classical vacuum is $Z_2$-symmetrical, the shape of the potential
$U(\phi)$ is mainly a single well, and when the vacuum is asymmetrical, the
shape of $U(\phi)$ is dominated by a double wells. These, according to the
above simple analysis, are independent of the coupling parameter $\beta$.
Moreover, the curvatures of the wells at their bottoms ${\frac {d^2 U(\phi)}
{(d \phi)^2}}\bigl |_{\phi=\pm\hat{\phi}_1}$ are also independent of $\beta$,
because for any given $m_0^2$ and $m^2$, any non-zero root of Eq.(2)
$\hat{\phi}_i$ is different for a different $\beta$ but the products
$\beta\hat{\phi}_i$s for all $i$ are the same with one another. Nevertheless,
$U(\hat{\phi_1})$ the depth of the $Z_2$-symmetry-broken wells with respect to
$U(\phi=0)$ is deeper and deeper with the decrease of $\beta$.

In this section, we have given a wordy statement on the $Z_2$-symmetry of the
classical vacuum. Next, we shall give the expression of the FTGEP so as to
discuss the $Z_2$-symmetry of quantum vacuum.

\section{Finite Temperature Gaussian Effective Potential}
\label{3}

The FTGEP is the effective potential of finite temperature field theory
\cite{33} obtained by the Gaussian approximation. There are several methods to
calculate the FTGEP \cite{35}, and those methods are equivalent to one another.
This section will calculate the FTGEP for the system Eq.(1) with the help of
the GWFA in thermofield dynamics \cite{35}(Roditi)\cite{36}.

Ref.~\cite{35}(Roditi) proposed the GWFA in thermofield dynamics for the
quantum-mechanical $\lambda\phi^4$ model. In Ref.~\cite{36}, we developed this
method for a (D+1)-dimensional scalar field theory with any potential whose
Fourier representation exists in the sense of tempered distributions \cite{38}.
According to the formulae in Ref.~\cite{36}, in order to calculate the FTGEP of
the 2D MSGFT, it is enough to complete some ordinary integrals because the
system Eq.(1) involves in the type of field systems in Ref.~\cite{36}. Hence,
it is not necessary to give the derivation of the FTGEP of the 2D MSGFT in
detail. Additionally, thermofield dynamics and its GWFA will be not introduced
here, for they can be found in Ref.~\cite{33,34} and in Ref.~\cite{36} ($C_1$
in Ref.~\cite{36} (page 744) should be $C_2$), respectively.

Now we shall give the FTGEP for the system Eq.(1). Substituting $U(\phi_x)$ of
Eq.(1) into Eqs.(23) and (25) in Ref.~\cite{36} and performing some
integrations, one can have the FTGEP of the 2D MSGFT at any temperature T
\begin{eqnarray}
V_T(\varphi)&=&{\frac {1}{2}}[J_0(g)-I_0(M^2)]
            -{\frac {1}{4}}[\mu^2 J_1(g)-M^2 I_1(M^2)]  \nonumber \\
            &\;& +{\frac {1}{4}}m_0^2[J_1(g)-I_1(M^2)]
            +{\frac {1}{2}}m_0^2\varphi^2     \nonumber \\
            &\;&  +{\frac {m^2}{\beta^2}}[1-\exp\{-{\frac {\beta^2}{4}}
            [J_{1}(g)-I_1(M^2)]\}\cos(\beta\varphi)] \nonumber \\
            &\;& -k_b T \int_{-\infty}^{\infty} {\frac {dp}{2 \pi}}
     [\cosh^2 (g(p,T))\ln(\cosh^2 (g(p,T)))
    - \sinh^2 (g(p,T))\ln(\sinh^2 (g(p,T)))] \; ,
\end{eqnarray}
with the gap equation
\begin{equation}
\mu^2(\varphi,T)= m_0^2+m^2 \exp\{-{\frac {\beta^2}{4}}
         [J_{1}(g)-I_1(M^2)]\}\cos(\beta\varphi) \; ,
\end{equation}
where $k_b$ represents the Boltzmann constant, $M$ is normal-ordering mass (an
arbitrary positive constant with mass dimension),
$$I_n[y^2]=\int_{-\infty}^{+\infty} {\frac {dp}
{2\pi}}{\frac {\sqrt{p^2+y^2}}{(p^2+y^2)^n}}$$
and $$J_n(g)=\int_{-\infty}^{+\infty} {\frac {dp \sqrt{p^2+\mu^2(\varphi,T)}
     \cosh(2g(p,T))}
{2 \pi (p^2+\mu^2(\varphi,T))^{ n}}}$$
with
$$g(p,T)={\frac {1}{2}}\ln\biggl({\frac {\exp\{{\frac {1}{2 k_b T}}
\sqrt{p^2+\mu^2(\varphi,T)}\}+1}
{\exp\{{\frac {1}{2 k_b T}}\sqrt{p^2+\mu^2(\varphi,T)}\}-1}}\biggr) \; .$$
Notice that in Eq.(5), $\mu^2(\varphi,T)$ in $J_n(g)$ should be replaced by
$\mu^2$.

In the above, $\varphi$ is the Gaussian-wave-functional expectation value of
the field operator \cite{36}, can be any real constant, and the quantun vacuum
is located at $\varphi=\varphi_0$ which is satisfied by the
equation
\begin{equation}
 m_0^2\varphi+ {\frac {m^2}{\beta}}\exp\{-{\frac {\beta^2}{4}}
         [J_{1}(g)-I_1(M^2)]\}\sin(\beta\varphi)=0
\end{equation}
and minimizes absolutely $V_T(\varphi)$ with respect to $\varphi$. Eq.(7) is
the extreme condition for $V_T(\varphi)$ and is obtained from the following
equation
\begin{equation}
{\frac {d V_T(\varphi)}{d \varphi}}=\int_{-\infty}^{+\infty}
    {\frac {d\alpha}{2\sqrt{\pi}}}e^{-{\frac {\alpha^2}{4}}}
    U^{(1)}({\frac {\alpha}{2}}\sqrt{J_1(g)- I_1 (M^2)}+\varphi) \; .
\end{equation}
Here, $U^{(1)}(y)\equiv {\frac {d U(y)}{dy}}$. If $\varphi_0\not= 0$, then the
symmetry $\phi\to -\phi$ is spontaneously broken at the quantum level and a
finite temperature.

The GWFA is a variational method. Because of the nature of the minimizing
procedure, $\mu$ in Eq.(5) should be chosen from the non-zero root of Eq.(6)
and two end points of the range $0<\mu<\infty$ so that $V_T(\varphi)$ is
an absolute minimum with respect to $\mu$. Besides, sometimes the non-zero
solution of Eq.(6) is multi-valued, and so in that case, the suitable root
should be decided according to the stability condition (from Eq.(26) in
Ref.~\cite{36})
\begin{equation}
1-{\frac {m^2\beta^2}{8}}\exp\{-{\frac {\beta^2}{4}}
         [J_{1}(g)-I_1(M^2)]\}J_2(g)\cos(\beta\varphi)
         >0 \;.
\end{equation}
Once $\mu$ is determined, its value at $\varphi_0$ is the thermal mass of
bosons Eq.(5) gives the FTGEP in the 2D MSGFT \cite{36}.

From Ref.~\cite{36}, the above formulae contain no divergences, and hence no
further renormalization procedures need to be performed. Thus, Eqs.(5),(6),(7)
and the inequality Eq.(9) can correctly give the FTGEP for the 2D MSGFT. From
the FTGEP, we can discuss the restoration and dynamical breakdown of the
$Z_2$-symmetry. Next, we begin in the $T=0$ case.

\section{Quantum Effects}
\label{4}

As was mentioned in Sect.I, for the system Eq.(1), with the aid of the GWFA,
Ref.~\cite{21} showed the occurence of the spontaneous $Z_2$-symmetry breakdown
and gave the corresponding phase diagram of the parameter plane
$\beta^2$-$m_0^2$. In this section, we shall discuss further the phases of the
quantum vacuum on the parameter planes $\beta^2$-$m^2$ and $m^2$-$m_0^2$ so as
to explicitly analyse the influence of quantum effects on the symmetry of
vacuum \footnote{It is difficult for Ref.~\cite{21} to explicitly discuss this
influence because the parameter $m^2$ was not considered in the numerical
calculation there.}.

When $T=0$, Eqs.(5), (6),(7) and the inquality (9) are reduced to the potential
\begin{equation}
\bar{V}_0(\varphi)
={\frac {\mu^2-M^2}{8\pi}}+{\frac {m_0^2+m^2-\mu^2}{\beta^2}}
  -{\frac {m_0^2}{8\pi}}\ln({\frac {\mu^2}{M^2}})
  +{\frac {1}{2}}m_0^2\varphi^2    \;,
\end{equation}
the gap equation
\begin{equation}
\mu^2=m_0^2+m^2({\frac {\mu^2}{M^2}})^
      {{\frac {\beta^2}{8\pi}}}\cos(\beta\varphi) \;,
\end{equation}
the extremum condition
\begin{equation}
m_0^2\varphi+{\frac {\mu^2-m_0^2}{\beta}}
         \tan(\beta\varphi)=0
\end{equation}
and the stability condition
\begin{equation}
1-\beta^2{\frac {\mu^2-m_0^2}{8\pi\mu^2}}
         >0 \;,
\end{equation}
which are consistent with Eqs.(4), (5), (6) and the inequality Eq.(7) in
Ref.~\cite{21}, respectively. From the discussion in Ref.~\cite{21}, the GEP
$V_0$ is governed by the nonzero root of Eq.(11), which is also satisfied by the
inequality, instead of the end points $\mu^2=0$ and $\mu^2\to \infty$, and
meanwhile, the coupling parameter $\beta^2$ is constrained to the range of
$0\le\beta^2<8\pi$. (Note that in Eqs.(10), (12) and (13), we have used
Eq.(11).)

Different from that choice in Ref.~\cite{21}, now letting $M^2=m_0^2$ and
further defining the following dimensionless quantities
\begin{equation}
{\bar{V}_0}(\varphi)\equiv {\frac {V_{T=0}(\varphi)}{m_0^2}},
{\bar{\mu}}\equiv {\frac {\mu}{m_0}},
{\bar{m}}\equiv {\frac {m}{m_0}} \;,
\end{equation}
one can numerically calculate the GEP at any fixed $m_0^2$ from Eqs.(10)---(14),
and accordingly analyse the phases of quantum vacuum on the parameter plane
$\bar{m}^2$-$\beta^2$. From the numerical results, the phase diagram of the
quantum vacuum (a figure which indicates the regions on the parameter space
where the corresponding vacua are $Z_2$-symmetrical and asymmetrical,
respectively) is depicted in the parameter plane $\bar{m}^2$-$\beta^2$ as Fig.4.
In this figure, the solid curve is the boundary of the symmetric and asymmetric
phases where the vacuum can be located either at $\varphi_0=0$ or at $\varphi_0
\not=0$, $i.e.$, at which the GEP at $\varphi_0=0$ is equal to the one at
$\varphi_0\not=0$. On the right of the boundary, the quantum vacuum is
$Z_2$-symmetrical, whereas on the left, the $Z_2$-symmetry of quantum vacuum is
spontaneously broken. The short-dashed line corresponds to the solid vertical
line in Fig.2(b). Thus, Fig.4 indicates that the symmetry $\phi\to -\phi$ is
restored by quantum effects in the domain I which is surrounded by the solid
curve and the short-dashed line. Obviously, for the parameter $\bar{m}^2$,
there is a critical value of $\bar{m}^2_c\approx -1.68$, and when $\bar{m}^2<
\bar{m}^2_c$ no $Z_2$-symmetric vacuum can appear. For any fixed $\bar{m}^2$
with $\bar{m}_c^2< \bar{m}^2<-1.0$, there are two critical values of $\beta$,
$\beta_{c1}$ and $\beta_{c2}$, at each of which the vacuum is either symmetrical
or asymmetrical, and for any $\beta$ with $\beta_{c1}<\beta <\beta_{c2}$, the
vacuum is located at $\varphi_0=0$. For an explicit illustration, we plot the
GEP in Fig.5 for the case of $\bar{m}^2=-1.4$. The curves II and IV correspond
to the two critical cases. Thus, one has seen that at any fixed $m_0^2$,
quantum effects can restore the $Z_2$-symmetry.

Perhaps one has noticed that there are only one critical value of $\beta$ for
the $\beta$-$\tilde{m}_0^2$ phase diagram in Ref.~\cite{21}, whereas now on the
plane $\beta$-$\bar{m}^2$ the two critical values $\beta_{c1}$ and $\beta_{c2}$
exist for any $\bar{m}^2$ with $\bar{m}_c^2< \bar{m}^2<-1.0$. This difference
between the $\beta$-$\bar{m}^2$ and $\beta$-$\tilde{m}_0^2$ diagrams can be
elucidated as per Fig.6. In Fig.6, the solid curve is the boundary between the
symmtric and asymmetric phases on the plane $\bar{\mu}^2_0$-$\beta^2$ (here,
$\bar{\mu}^2_0={\frac {\mu^2(\varphi=0)}{m_0^2}}=\tilde{m}^{-2}_0$), which
corresponds to the long-dashed boundary in the parameter plane
$\beta^2$-$\tilde{m}^2_0$ of Fig.1 in Ref.~\cite{21}, and the short- or
tiny-dashed curves I, II, III, IV and V are plotted from the expression of
$\bar{\mu}^2_0$ for the given values of $\bar{m}$: $-2.0, \bar{m}^2_c\approx
-1.68, -1.3, -1.001$ and $-0.8$, respectively. Fig.6 indicates that in the
parameter plane $\bar{\mu}_0^2$-$\beta^2$, any curve $\bar{m}^2=constant$ with
$\bar{m}_c^2< \bar{m}^2<-1.0$, such as the curve III, intersects the solid
curve at two points which correspond to $\beta_{c1}$ and $\beta_{c2}$.

In the above, the $Z_2$DSB phenomenon is not displayed on the parameter plane
$\bar{m}^2$-$\beta$. Nevertheless, we dare not rashly have a conclusion that no
$Z_2$DSB can occur for the 2D MSGFT, because this theory has three parameters
and the parameter plane $m_0^2$-$m^2$ is not considered still. Now, we turn on
it. Letting the normal-ordering mass unit, and giving an explicit value of
$\beta$, we can discuss the phase diagram of quantum vacuum on the parameter
plane $m_0^2$-$m^2$. For the case of $\beta^2=3.0$, we numerically solved
Eqs.(10)---(13) (when $M$ is unit, all quantities can be regared as
dimensionless ones), and the phase diagram on the parameter plane $m_0^2$-$m^2$
is plotted in Fig.7 (only one part of the fourth quadrant is given). In Fig.7,
the solid curve is the boundary between the symmetric and asymmetric phases,
and the region above the solid curve corresponds to symmetric vacua and the
region below the solid curve to asymmetric vacua. The short-dashed line in this
figure is the boundary in Fig.2(a), and obviously the figure indicates that not
only the $Z_2$-symmetry restoration occurs in the upper-left region I between
the solid curve and the short-dashed line, but also the $Z_2$-symetry is
dynamically spontaneously broken in the lower-right region II between the solid
curve and the short-dashed line. In order to give an evident demonstration for
the restoration and dynamical breakdown of the $Z_2$-symmetry, the GEP is
depicted in Fig.8 and Fig.9, respectively. Fig.8 is for the case of $m_0^2=1.5$,
and the curves I, II and III correspond to $m^2=-1.7, -2.1526852$ (the
approximate critical value) and $-3.0$, respectively. In Fig.9, $m_0^2=8.0$, and
the curves I, II and III correspond to $m^2=-5.0, -6.3043705$ (the approximate
critical value) and $-7.0$, respectively. The dynamical breakdown of the
$Z_2$-symmetry in the 2D MSGFT is consistent with assertion of Ref.~\cite{1}
(notice the third footnote. In the paragraph above Fig.2 of Ref.~\cite{21}, the
second sentence ``Fr\"ohlich [1] pointed out $\cdots$'' is misleading, and
should be ``Fr\"ohlich [1](the book) pointed out that for a sufficiently small
$\tilde{m}_0^2$ there may be a phase transition. ''). Additionally, we
also consider the effect of the parameter $\beta^2$ on the phase diagram of the
plane $m^2$-$m^2_0$, and the result indicates that the intersection point of
the solid boundary with the short-dashed boundary is shifted up along the
classical boundary with the increase of $\beta^2$. That is to say, when
increasing the value of $\beta^2$, the $Z_2$-symmetry restoration domain I gets
smaller and the $Z_2$DSB domain II larger. As a illustration, in Fig.7, we also
plot the boundaries in the cases of $\beta=2.4$ (the lower tiny-dashed curve)
and $\beta=4.3$ (the upper tiny-dashed curve).

In this section, we have explicitly demonstrated the existence of the
restoration and the dynamical breakdown of the $Z_2$-symmetry in the 2D MSGFT.
Next section, we shall analyse the FTGEP to discuss the influence of a finite
temperature effect on the results in this section.

\section{Finite Temperature Effects}
\label{5}

Substituting the three equations between Eqs.(26) and (27) in Ref.~\cite{36}
into Eqs.(5)---(7),(9) here, we have the FTGEP
\begin{eqnarray}
V_T(\varphi) &=& {\frac {\mu^2-M^2}{8\pi}}+{\frac {1}{2}}m_0^2\varphi^2
  -{\frac {m_0^2}{8\pi}}\ln({\frac {\mu^2}{M^2}}) \nonumber \\
  & \ \ \ & {\frac {m^2}{\beta^2}}[1-({\frac {\mu^2}{M^2}})^
   {{\frac {\beta^2}{8\pi}}}\exp\{-{\frac {1}{2}}\beta^2 C_2\}
   \cos(\beta\varphi)] +{\frac {1}{2}} C_2 (m_0^2-\mu^2) +k_b T K  \;,
\end{eqnarray}
the gap equation
\begin{equation}
\mu^2=m_0^2+m^2({\frac {\mu^2}{M^2}})^{{\frac {\beta^2}{8\pi}}}
      \exp\{-{\frac {1}{2}}\beta^2 C_2\}\cos(\beta\varphi) \;,
\end{equation}
the extremum condition
\begin{equation}
  m_0^2\varphi+{\frac {m^2}{\beta}}
  ({\frac {\mu^2}{M^2}})^{{\frac {\beta^2}{8\pi}}}
      \exp\{-{\frac {1}{2}}\beta^2 C_2\}\sin(\beta\varphi)=0
\end{equation}
and the stability condition
\begin{equation}
1- {\frac {1}{8}}m^2\beta^2({\frac {\mu^2}{M^2}})^{{\frac {\beta^2}{8\pi}}}
      \exp\{-{\frac {1}{2}}\beta^2 C_2\}J_2(g)\cos(\beta\varphi)
         >0 \;,
\end{equation}
where
$$C_2=\int_{-\infty}^{+\infty}{\frac {d p}{2 \pi}}
   {\frac {1}{\sqrt{p^2+\mu^2}[\exp\{{\frac {1}{k_b T}}
   \sqrt{p^2+\mu^2}\}-1]}}$$
and
$$K=\int_{-\infty}^{+\infty}{\frac {d p}{2 \pi}}
      \ln[1-\exp\{-{\frac {1}{k_b T}}\sqrt{p^2+\mu^2}\}].$$
Note that $\mu$ in Eqs.(16)---(18) represents $\mu(\varphi,T)$.
From the formulae 3.337(1) and 8.526(1) in Ref.~\cite{39}, one can have
$$C_2={\frac {\gamma}{2\pi}}+{\frac {1}{2\pi}}\ln({\frac {\mu}{4\pi k_b T}})
      + {\frac {k_b T}{2\mu}}+{\frac {1}{2\pi}}\sum_{n=1}^{\infty}
    [{\frac {2\pi k_b T}{\sqrt{\mu^2+4\pi^2 k_b^2 T^2 n^2}}}-{\frac {1}{n}}]$$
with $\gamma$ the Euler's constant. According to Sect.III, in order to
calculate the FTGEP, we have to first choose $\mu$ from the root of the gap
equation Eq.(16) and the end points of the region $0\le\mu<\infty$. For the end
point $\mu\to 0$, the integral $K$ is finite,
$C_2\to {\frac {1}{2\pi}}\ln({\frac {\beta\mu}{4\pi}})
+{\frac {1}{2\beta\mu}}\to {\frac {1}{2\beta\mu}}$, and so $V_T\to
{\frac {m_0^2 k_b T}{4 \mu}}\to +\infty$. Therefore, we have to discard $\mu=0$.
As for the other end point $\mu\to \infty$, $K=0$ and $C_2=0$, and hence
$$V_T(\varphi)\to {\frac {\mu^2}{8\pi}}-{\frac {m^2}{\beta^2}}
    ({\frac {\mu^2}{M^2}})^{\frac {\beta^2}{8\pi}}cos(\beta\varphi) ,$$
which is the same as that in the case of $T=0$ \cite{21}. Thus, $\beta^2$ must
be less than $8\pi$, and to calculate FTGEP, we should choose the nonzero root
of Eq.(16) instead of the end points $\mu=0$ and $\mu\to \infty$ (Of course,
the root should be satisfied by the inequality Eq.(18)). The constrant of the
parameters in the case of $T\not=0$ is identical with that in the case of $T=0$.

Now we can compute the FTGEP and analyse the symmetry of the vacuum at
$T\not=0$. Although the integrals $K$, $C_2$ and $J_2(g)$ have no simple
analytical expressions, but it is easy to numerically compute them, and
accordingly we can numerically solve the Eqs.(15)---(18). For the convenience
of a comparison, we shall do as the last section and consider first the phase
diagram on the parameter plane $m^2$-$\beta^2$ and then that on the plane
$m_0^2$-$m^2$.

Taking $M=m_0$, using the definitions Eq.(14) and defining the dimensionless
FTGEP
\begin{equation}
{\bar{V}_T}(\varphi)\equiv {\frac {V_{T}(\varphi)}{m_0^2}} \;,
\end{equation}
Eqs.(15)---(18) can be dimensionless, and we can numerically computed the FTGEP
at a given temperature and discuss the symmetry of the vacuum on the parameter
plane $\bar{m}^2$-$\beta^2$. Fig.10 gives the numerical results at some
temperatures. In Fig.10, the solid curves I,II and III are the boundaries
between symmetric and asymmetric phases at $T={\frac {1}{10 k_b}},
{\frac {1}{2 k_b}}$ and ${\frac {1}{k_b}}$, respectively, the shapes of which
are similar to that in Fig.4, the short-dashed line corresponds to the classical
boundary in Fig.2(b), and the curve I coincides almost with the boundary in
Fig.4. At any temperature $T$, the vacuum is asymmetrical on the left of the
solid curve of Fig.10 and symmetrical on the right. Fig.10 indicates that a
finite temperature effect can further restore $Z_2$-symmetry of the quantum
vacuum, and the $Z_2$-symmetry restoration domain is the domain between the
curves II and I for the case of $T={\frac {1}{2 k_b}}$ and the one between the
curves III and I for the case of $T={\frac {1}{k_b}}$. It is evident that with
the increase of the temperature the $Z_2$-symmetry restoration domain get larger
and larger. As an illustration of the $Z_2$-symmetry restoration, for the case
of $\bar{m}^2=-1.4$ and $T={\frac {1}{2 k_b}}$, the FTGEPs at $\beta=4.71,
\beta_{c2}\approx 4.635, 3.0, \beta_{c1}\approx 1.105$ and $0.85$ are plotted
as the curves I,II,III,IV and V in Fig.11, respectively.

Now we are in a position to consider the parameter plane $m_0^2$-$m^2$. Letting
$M$ unit, quantities in Eqs.(15)---(18) can be regarded as dimensionless, and
from Eqs.(15)---(18), we can compute numerically the FTGEP at a given
temperature and a given $\beta$ for discussing the phases of the vacuum on the
plane $m_0^2$-$m^2$. Numerical analyses indicate that a finite temperature can
restore further the $Z_2$-symmetry still, but cannot give rise to $Z_2$DSB. At
$\beta=3.0$, the phase diagrams for some temperatures are plotted in Fig.12,
and the phase diagrams at any other $\beta$ are similar to Fig.12. In Fig.12,
the solid curve is the boundary between the symmetric and asymmetric phases for
$T={\frac {1}{2 k_b}}$, at any point of which $V_T(\varphi=\varphi_0=0)$ is equal
to $V_T(\varphi=\varphi_0\not=0)$ and the vacuum is either symmetrical or
asymmetrical. For the region above the boundary, the vacuum is symmetrical,
whereas for the region below the boundary the vacuum is asymmetrical. The upper
and lower tiny-dashed curves are the boundaries for $T={\frac {1}{6 k_b}}$ and
${\frac {1}{k_b}}$, respectively, and the upper tiny-dashed curve coincide
almost with the corresponding boundary at $T=0$, $i.e.$, the solid curve in
Fig.7. For a sufficient large $m_0^2$, the boundary at any $T\not=0$ coincides
presumably with the zero-temperature boundary \footnote{Up to $m_0^2=30.0$ the
numerical computation showed such a coincidence. Here we use the word
``presumably'' because it is impossible to check numerically all values of
$m_0^2$}. Fig.12 indicates that the finite-temperature effect compresses the
$Z_2$DSB domain at $T=0$ and enlarge the $Z_2$-symmetry restoration domain at
$T=0$. Thus, the finite-temperature effect leads to no $Z_2$DSB but a further
restoration of the $Z_2$-symmetry. It is evident from Fig.12 that the higher
the temperature is, the larger the $Z_2$-symmetry restoration domain is. For an
illustration of the phase diagram and the $Z_2$-symmetry restoration by the
finite temperature effect, Fig.13 and Fig.14 give the FTGEPs at $\beta=3.0$ for
some cases. In Fig.13, the curves I, II and III are the FTGEPs at
$T={\frac {1}{2 k_b}}$ and $m_0^2=1.5$ in the cases of the $m^2=-1.7,
-2.4821738$ and $3.0$, respectively, and meanwhile for the third case we also
give the FTGEP at $T={\frac {1}{k_b}}$ as the curve IV. In Fig.14, the curves I,
II and III are the FTGEPs at $T={\frac {1}{2 k_b}}$ and $m_0^2=8.0$ in the cases
of $m^2=-5.0, -6.391072$ and $6.7$, respectively, and meanwhile for the third
case we also give the FTGEP at $T={\frac {1}{k_b}}$. It is noticed that the
points \{$m_0^2=1.5,m^2=-2.4821738$\} and \{$m_0^2=8.0,m^2=-6.391072$\} on the
parameter plane $m_0^2$-$m^2$ are located at the solid curve of Fig.12, and the
points \{$m_0^2=1.5,m^2=-3.0$\} and \{$m_0^2=8.0,m^2=-6.7$\} on the parameter
plane $m_0^2$-$m^2$ are located in the domain between the solid and the lower
tiny-dashed curves of Fig.12. Thus Fig.13 and Fig.14 plainly display the above
results.

\section{Conclusion}
\label{6}

The $Z_2$-symmetry of the vacuum of the 2D MSGFT was investigated both in
Ref.~\cite{21} and this paper. Ref.~\cite{21} gave the phase diagram on the
parameter plane $\tilde{m}_0^2$-$\beta^2$ for the zero-temperature case and
showed the existence of $Z_2$SSB, whereas the present paper further gave the
phase diagrams on the parameter planes $\bar{m}^2$-$\beta^2$ and $m_0^2$-$m^2$
both for $T=0$ and $T\not=0$ cases and exhibited the influences of quantum and
finite temperature effects on the symmetry of the vacuum. From this paper, the
2D MSGFT at the quantum level can suffer not only the restoration of
$Z_2$-symmetry but also $Z_2$DSB, and a finite temperature effect can enlarge
the $Z_2$-symmetry restoration phenomenon and compress the $Z_2$DSB, namely, a
finite temperature effect can further restore the $Z_2$-symmetry only.
Those conclusions in the present paper are interesting and enhance our
understanding on the 2D MSGFT. As was pointed out in the introduction, the GWFA
is believable, at least, qualitatively, albeit it is a simple approximation. In
fact, our results have confirmed and realized the inferences or predictions in
Refs.~\cite{1,15,16}. Although an investigation with the help of some better
approximate methods \cite{40} would perhaps have different numerical results
about the 2D MSGFT, the results in this paper will provide a basis and a
reference at least, for to our knowledge, no quantitative results about the
restoration and dynamical breakdown of $Z_2$-symmetry in the 2D MSGFT existed
in the literature.

Perhaps some results in this paper can be conceivable from quantum mechanics
\footnote{The behaviours in quantum field theory have a bit of similarity to
those in quantum mechanical case.}. Quantum effects altering the $Z_2$-symmetry
of the vacuum consists in that the differences of calssical potential at
different well-bottums (and/or field points) can be offsetted by the
corresponding differences of quantum correstions of the potentials \cite{22,24}.
Obviously, occurrences of both the restoration and the dynamical breakdown of
$Z_2$-symmetry have much to do with the depths and the bottum curvatures of the
wells of the classical potential (from quantum mechanics \cite{41}, the quantum
correstion to the classical energy gets more remarkable with the increase of
the curvature of the bottm). From Eqs.(3) and (4), the curvatures of the
well-bottums are governed by the parameters $m_0^2$ and $m^2$, and the depths
of the wells by the parameters $\beta$, $m_0^2$ and $m^2$. Thus, the concrete
results in this paper should be acceptable. In particular, some of them can be
understood simply from the above statement. For example, one can have a
straightforward understanding about the existence of $\beta_{c1}$ because from
Sect.II the depth of the $Z_2$-symmetry-broken well gets deeper and deeper with
the decrease of $\beta$.

Those figures about parameter planes and effective potentials in this paper
indicated the existence of some phase transitions. In Ref.~\cite{21}, we have
discussed the phase transition releted to the parameters $\beta$ and $m_0$,
at $T=0$, and our viewpoints about the phase transitions related to the present
cases resemble those in Ref.~\cite{21}. So here we discuss them no longer.
Nevertheless, we want to mention the meaning of $Z_2$-symmetry in the massive
Schwinger model. From the results here, the restoration and dynamical breakdown
of the $Z_2$-symmetry is relevant to the negative $m^2$. This reminds us that
the system Eq.(1) with any negative $m^2$ and $\beta^2=4\pi$ is equivalent to
the massive Schwinger model with $\pi$ vacuum angle and on the zero-charge
sector \footnote{In fact, when a statement in this paper is related to the
comparison between the results here and the inference in Ref.~\cite{15}, we have
used this equivalence.}. It is known that the symmetry $\phi\to -\phi$ is
charge conjugation in Fermi language, and the symmetric vacuum in the 2D MSGFT
corresponds to the disappearance of half-asymptotic particles (quarks and
antiquarks) in the massive Schwinger model and the asymmetric vacuum to the
occurence of half-asymptotic particles in the massive Schwinger model \cite{15}.
Hence, the restoration of $Z_2$-symmetry here implies a phase transition from a
half-asymptotic-particle phase to a no-quark phase, and the $Z_2$DSB
corresponds to a inverse phase transition, $i.e.,$ a phase transition from a
no-quark phase to a half-asymptotic-particle phase. This is an intereting
phenomenon and worthy of a detailed investigation.

Finally, we also to mension that it is very difficult to extract a
non-perturbative information when a finite-temperature effect is introduced
into a quantum field theory. Nevertheless, Ref.~\cite{35,36} and the present
paper has indicated that the GWFA is a viable and effective tool for
extracting the non-perturbative information of finite temperature field
theories. With the aid of the GWFA, we have calculated the masses of the
Schwinger boson and its two-particle bound states in Ref.~\cite{21}. From this
paper and Ref.~\cite{36}, after a finite temperature effect is introduced into
the massive Schwinger model, it will be still possible to calculate the masses
of the Schwinger boson and its bound states. We believe that the GWFA will
become an important tool in finite temperature field theory.

\acknowledgments
This project was supported jointly by the President Foundation of Shanghai Jiao
Tong University, and the National Natural Science Foundation of China with
grant No. 19875034.

\figure{Fig.1 A graphical solution of Eq.(2) with $\beta=3.0$. The lines I,II
        and III are $Y={\frac {m_0^2}{m^2}}\beta\phi$ with ${\frac {m_0^2}{m^2}}
        =-1.2, -0.5$, and $-0.1$, respectively. This figure indicates that for
        the three cases there are 1, 3 and 7 roots of Eq.(2), respectively.}
\figure{Fig.2 The parameter space of the classical 2D MSGFT corresponding to
        the symmetric and the asymmetric vacua. Fig.2(a) is on the
        the parameter plane $\{m_0^2$-$m^2\}$ at any given $\beta$, and Fig.2(b)
        is on the parameter plane $\{m^2,\beta^2\}$ at a given $m_0^2$.}
\figure{Fig.3 Classical potentials at $\beta=3.0$ and $m_0^2=0.6$ for some
        $m^2$s. The curves I, II, III and IV are for the cases of $m^2=6.0,
        -0.5, -3.0$ and $-6.0$, respectively.}
\figure{Fig.4 The phase diagram on the parameter plane $\bar{m}^2$-$\beta^2$ at
        $T=0$, which is plotted from Eqs.(10)---(14). The short-dashed line
        corresponds to the boundary in Fig.2(b), and the domain between it and
        the solid curve (the domain I) is the $Z_2$-symmetry restoration region.
\figure{Fig.5 The GEP of the 2D MSGFT at points on the parameter plane
        $\bar{m}^2$-$\beta^2$ in the case of $\bar{m}^2=-1.4$. Only one-half of
        the symmetric GEP is shown. In this figure, the curves I, II,
        III, IV and V correspond to $\beta^2=4.6, \beta_{c2}\approx 4.4685,
        3.0, \beta_{c1}\approx 1.8685$ and $1.3$, respectively.}
\figure{Fig.6 The elucidation of the existences of $\beta_{c1}$ and
        $\beta_{c2}$. The solid curve is transformed from the long-dashed curve
        in Fig.1 of Ref.~\cite{21}. The short- or tiny-dashed curves I,II,III,
        IV and V are plotted from the expression of $\bar{\mu}^2_0$ for the
        given values of $\bar{m}$: $-2.0, \bar{m}^2_c\approx -1.68, -1.3,
        -1.001$ and $-0.8$, respectively.}
\figure{Fig.7 The phase diagram on the parameter plane $m_0^2$-$m^2$ at $T=0$,
        which is plotted from Eqs.(10)---(13). In this figure, the solid curve
        is the boundary for $\beta=3.0$, the short-dashed line is the boundary
        in Fig.2(a), and the domains I and II are surrounded by the solid and
        short-dashed curves. Additionally, the upper and lower tiny-dashed
        curves are the other two boundaries in the cases of $\beta=4.3$ and
        $2.4$, respectively.
\figure{Fig.8 The GEP of the 2D MSGFT for the case of $m_0^2=1.5$ and
        $\beta=3.0$. Only one-half of the symmetric potential is shown.}
\figure{Fig.9  Similar to Fig.8 but for the case of $m_0^2=8.0$ and
        $\beta=3.0$.}
\figure{Fig.10  The $Z_2$-symmetry restoration on the parameter plane
        $\bar{m}^2$-$\beta$ at finite temperatures. The solid curves I,II and
        III are the boundaries between the symmetric and asymmetric phases for
        the temperatures $T={\frac {1}{10 k_b}}, {\frac {1}{2 k_b}}$ and
        ${\frac {1}{k_b}}$, repectively. The curve I coincides almost with the
        solid curve in Fig.4, and the short-dashed line corresponds to the
        boundary at the classical case in Fig.2(b).}
\figure{Fig.11 The FTGEPs of the 2D MSGFT with $T={\frac {1}{2 k_b}}$ at
        points on the parameter plane $\bar{m}^2$-$\beta^2$. In this figure,
        $\bar{m}^2=-1.4$ and the points \{-1.4,4.635\} and \{-1.4,1.105\} are
        located at the curve II in Fig.10.}
\figure{Fig.12 The restoration and dynamical breakdown of $Z_2$-symmetry on
        the parameter plane $m_0^2$-$m^2$ for the 2D MSGFT at finite
        temperatures. The solid curve is the boundary between the symmetric and
        asymmetric phases for $T={\frac {1}{2 k_b}}$, and the upper and lower
        tiny-dashed curves are the boundaries for $T={\frac {1}{6 k_b}}$ and
        ${\frac {1}{k_b}}$. The three boundaries are for the case of
        $\beta=3.0$. The short-dashed line is the boundary at the classical
        case in Fig.2(a), and the upper tiny-dashed curve coincides almost with
        the solid boundary in Fig.7. Besides, the domains I and II are those
        between the solid and short-dashed boundaries.}
\figure{Fig.13 The FTGEPs of the 2D MSGFT with finite temperatures at
        points on the parameter plane $m_0^2$-$m^2$. All curves are plotted at
        $m_0^2=1.5$, the curves I, II and III are for $m^2=-1.7, -2.4821738$
        and $-3.0$ at $T={\frac {1}{2 k_b}}$, respectively, and the curve IV
        corresponds to same point as the cuvre III but at $T={\frac {1}{k_b}}$.
        The point the curve II corresponds to is located at the solid boundary
        in Fig.12.}
\figure{Fig.14 Similar to Fig.13 but All curves are plotted at
        $m_0^2=8.0$. The curves I, II and III are for $m^2=-5.0, -6.391072$
        and $-6.7$ at $T={\frac {1}{2 k_b}}$, respectively, and the curve IV
        corresponds to same point as the cuvre III but at $T={\frac {1}{k_b}}$.
        The point the curve II corresponds to is located at the solid boundary
        in Fig.12.}

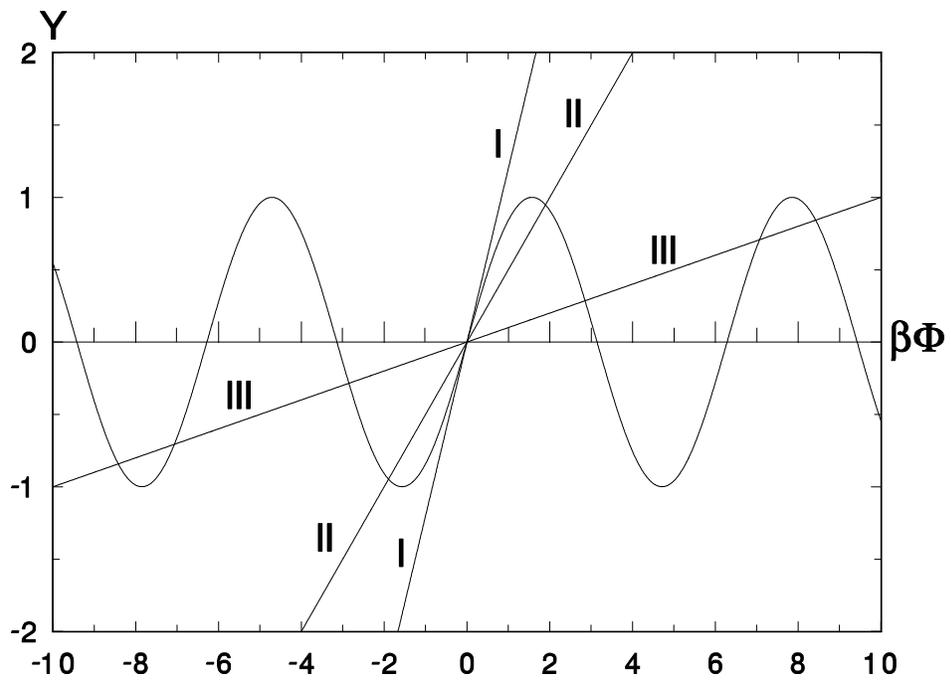
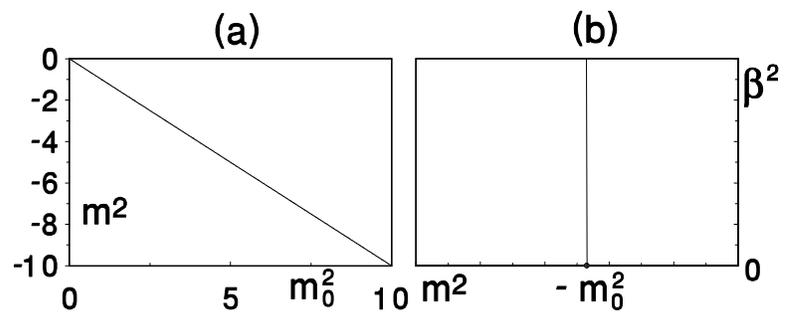
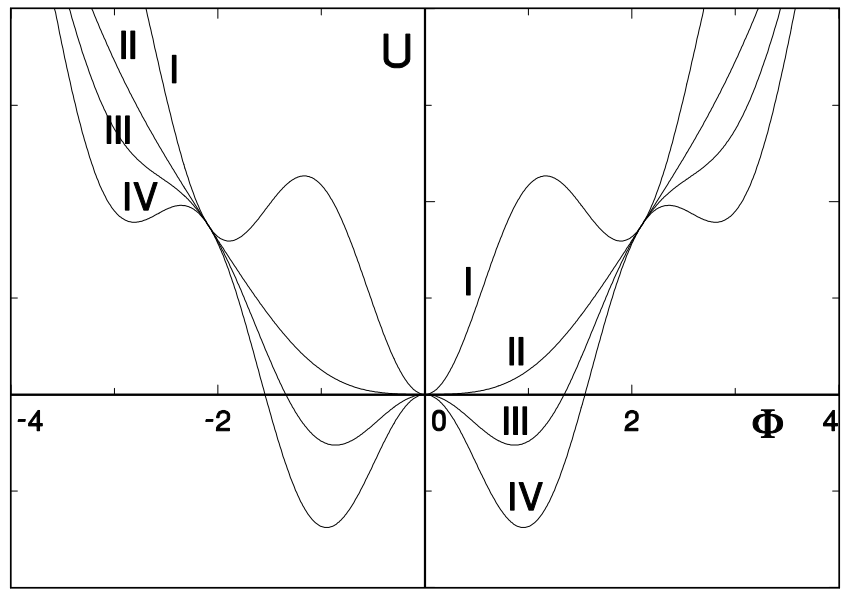
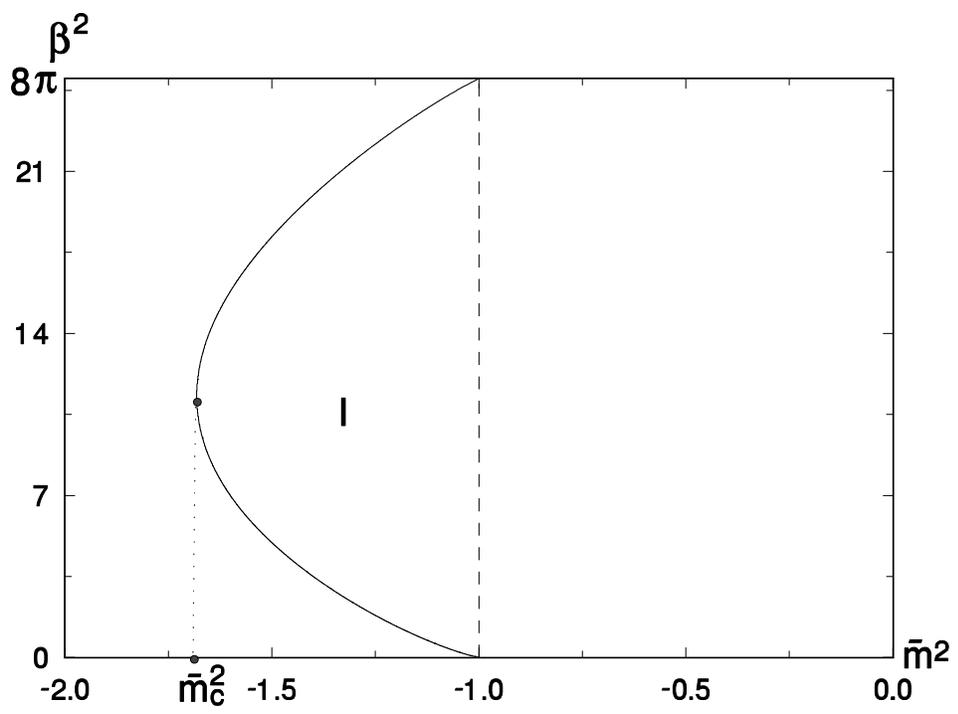
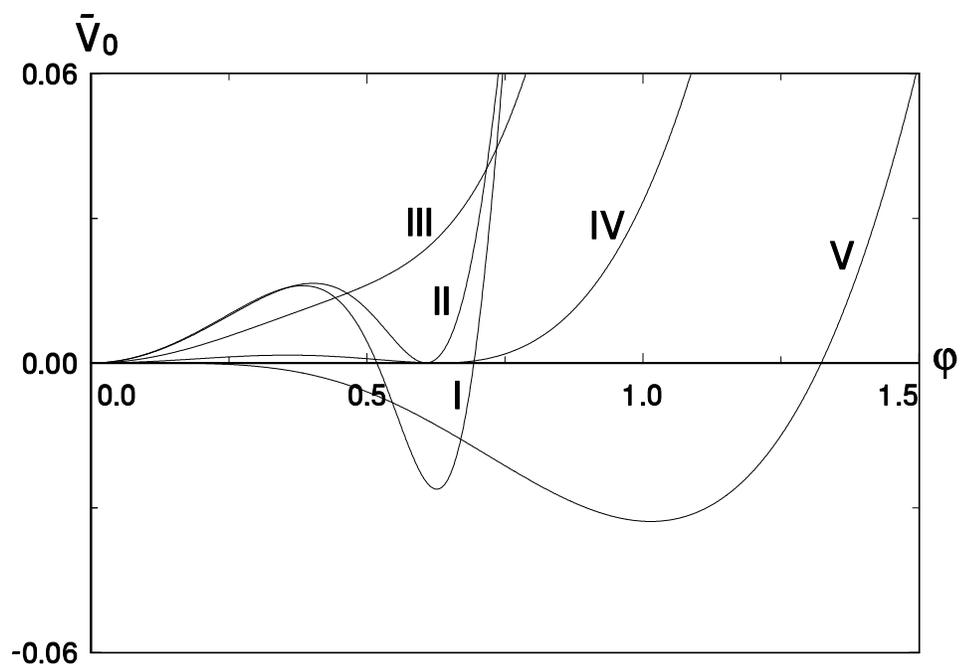
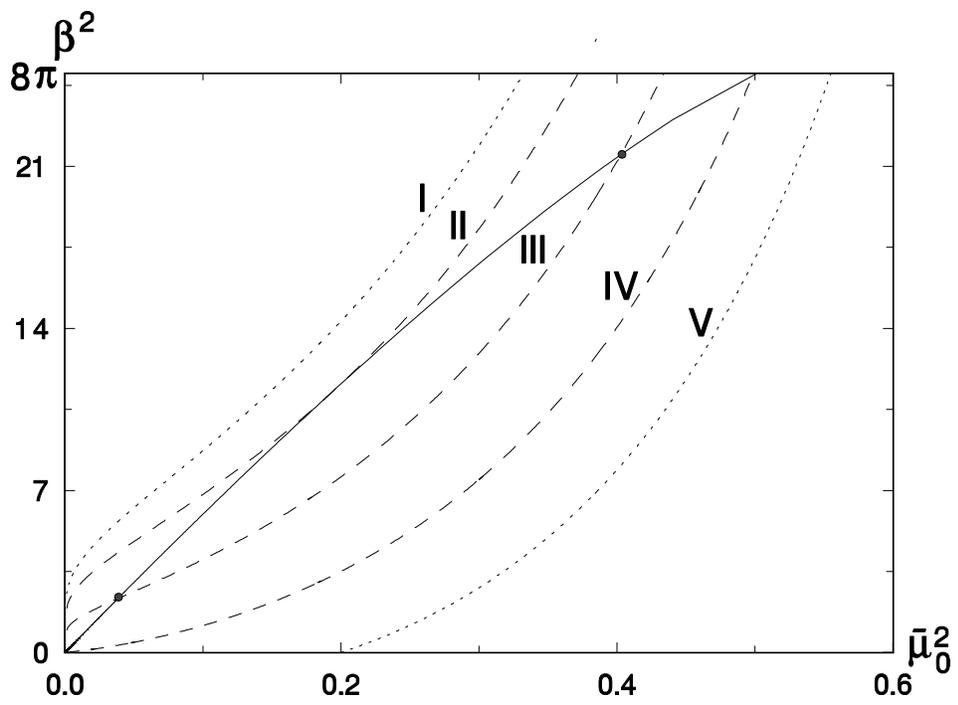
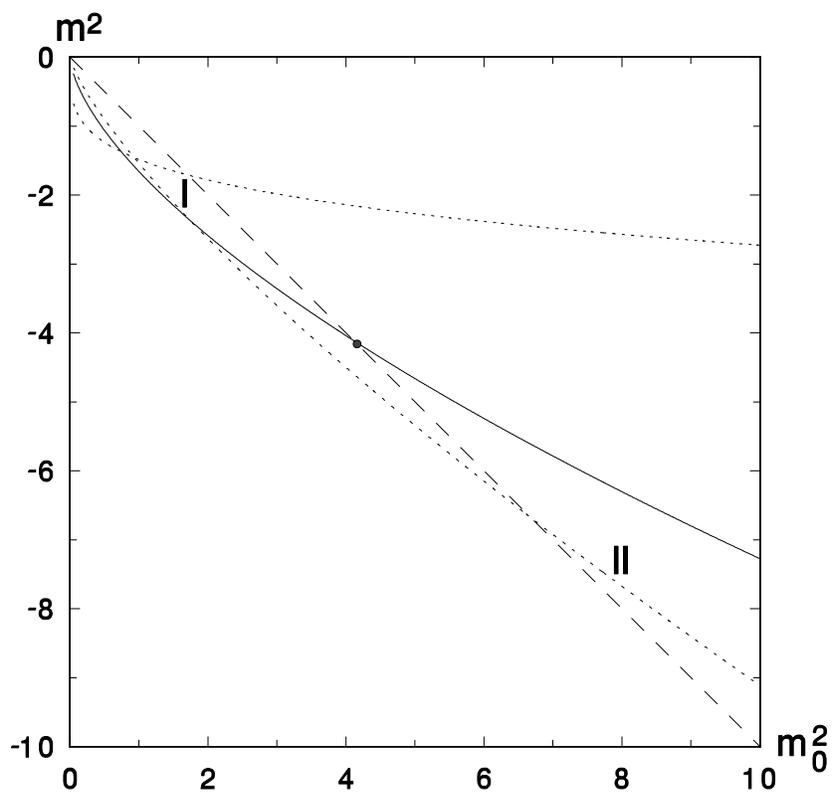
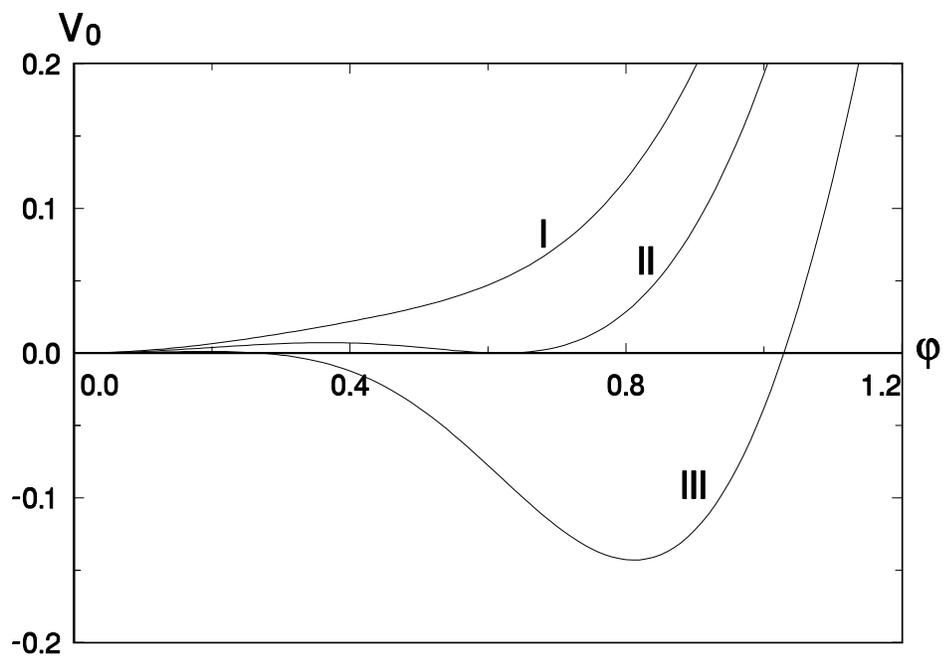
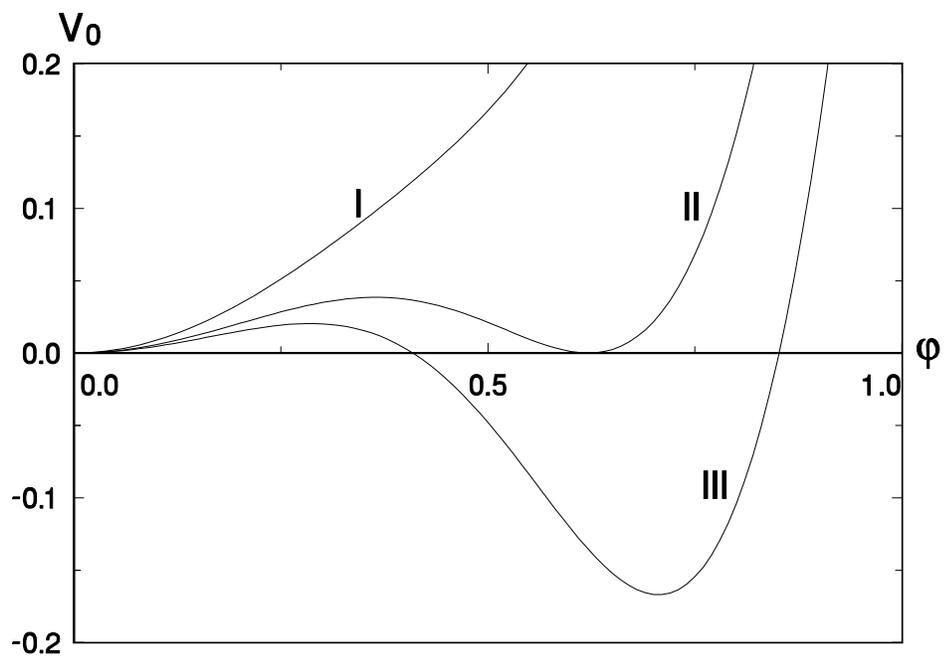
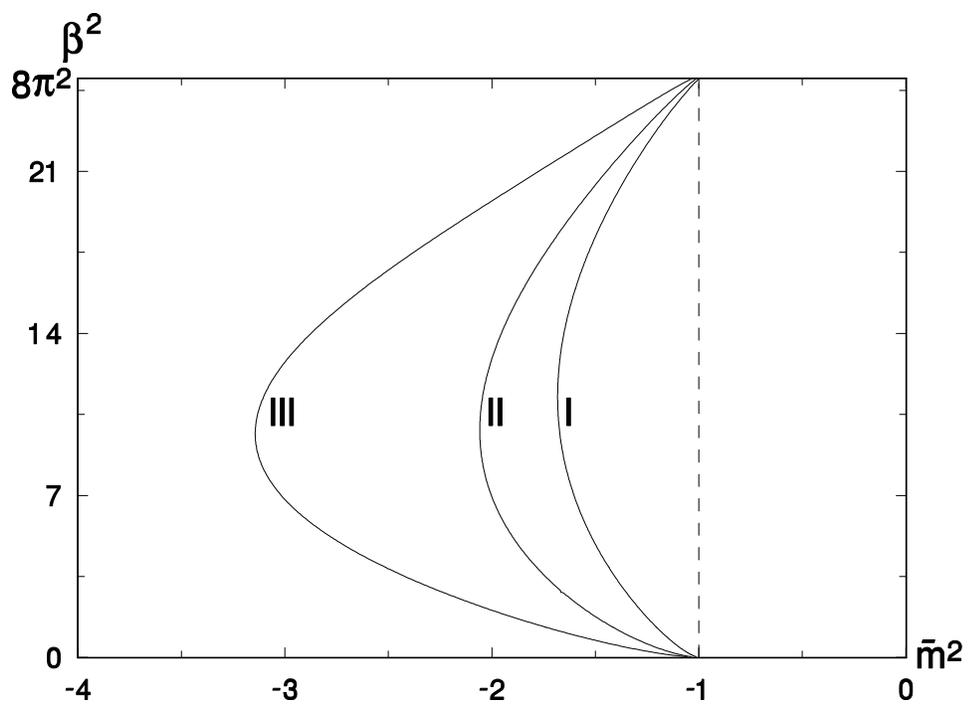
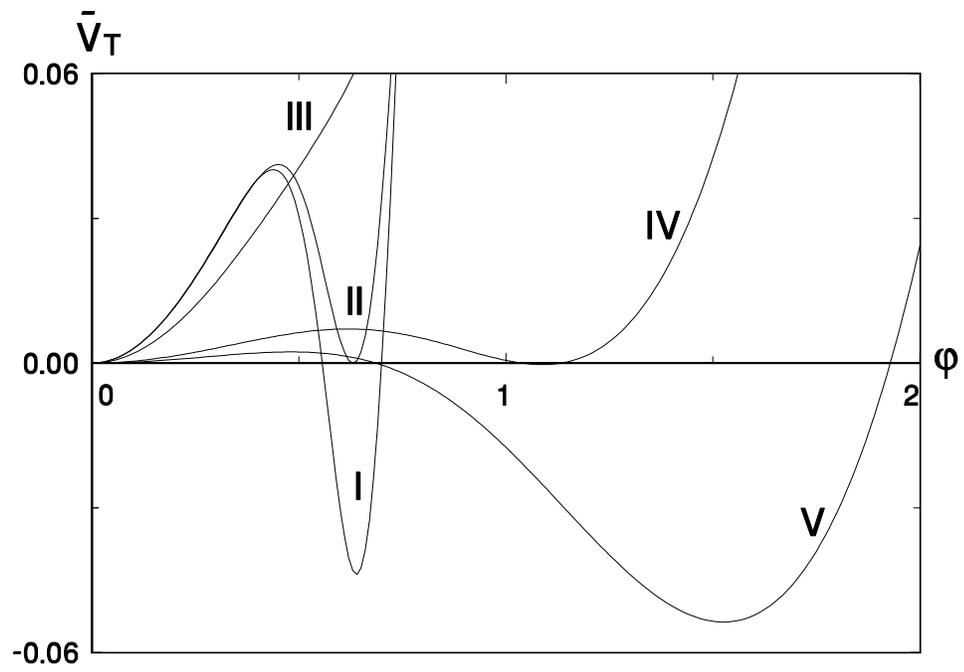
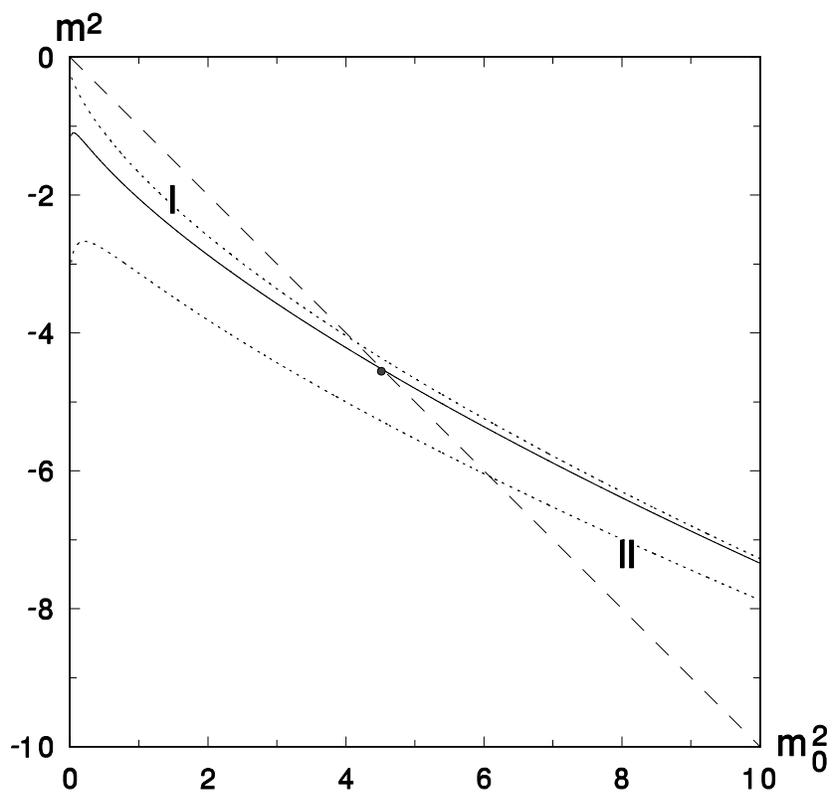
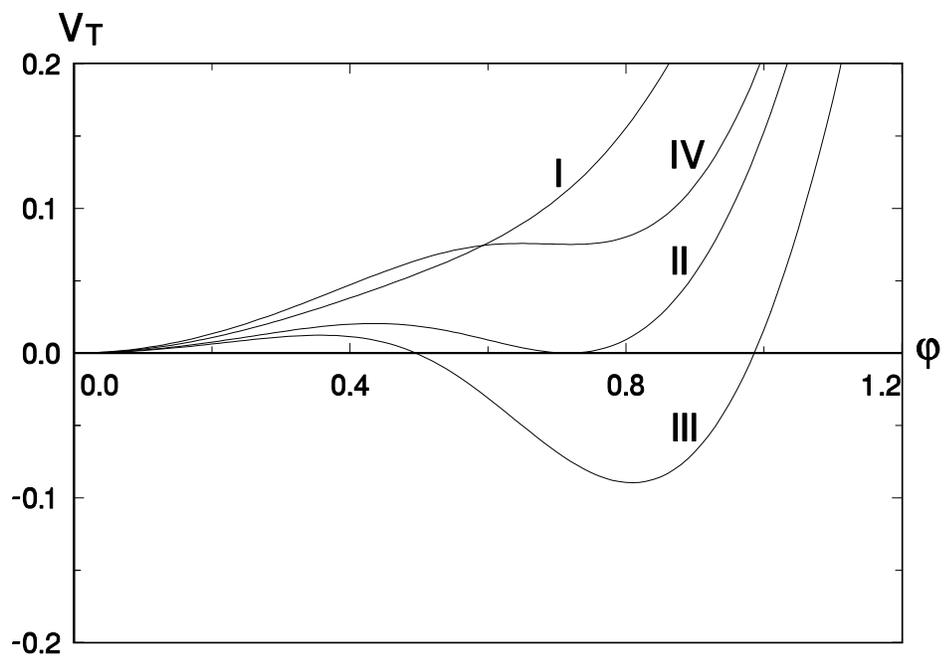
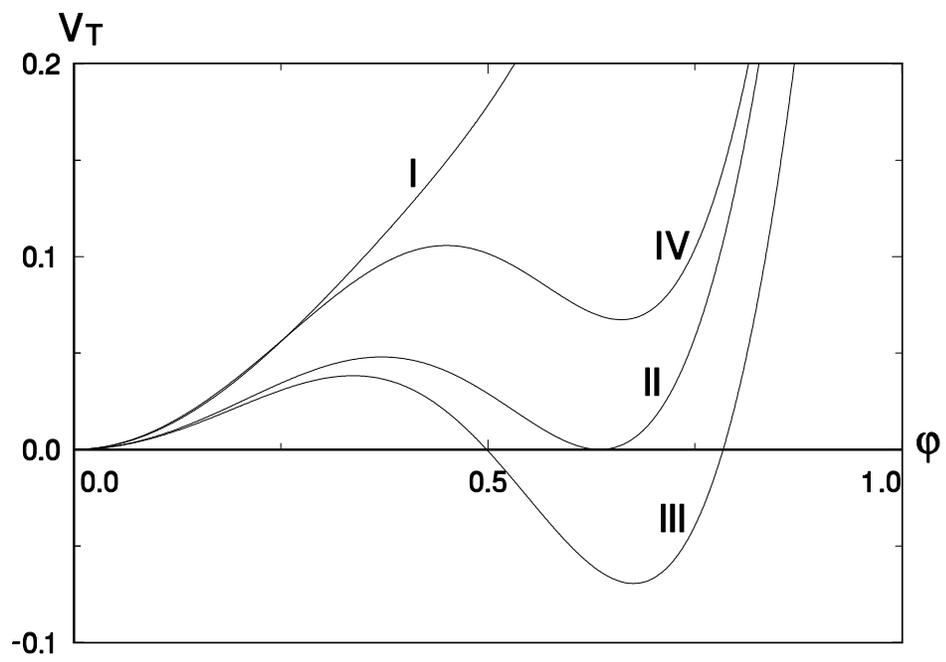
\end{document}